\documentclass[english]{article}
\usepackage[T1]{fontenc}
\usepackage[latin9]{inputenc}
\usepackage{units}
\usepackage{textcomp}
\usepackage{amstext}
\usepackage{setspace}
\usepackage{amssymb}
\usepackage{esint}

\makeatletter

\newcommand{\lyxmathsym}[1]{\ifmmode\begingroup\def\b@ld{bold}
  \text{\ifx\math@version\b@ld\bfseries\fi#1}\endgroup\else#1\fi}

\newcommand{\lyxaddress}[1]{
\par {\raggedright #1
\vspace{1.4em}
\noindent\par}
}

\makeatother

\usepackage{babel}

\begin{document}

\title{Interesting, Surprising and perhaps Amusing Results in Nuclear Physics
Calculations over the years}

\author{Compiled by Justin Farischon and Larry Zamick}

\maketitle

\lyxaddress{Department of Physics and Astronomy, Rutgers University Piscataway,
New Jersey, 08854 }
\begin{abstract}
We have compiled some sectors of works by L. Zamick and collaborators
which we hope will be of interest to the reader. Some of the results
which we, at least, have found amusing often popped up unexpectedly
in what were apparent routine calculations. We feel that the sections
we have made can be of importance in farthing the field but our main
criteria are that the results will have some educational value and
that they add a bit of zest to our field.
\end{abstract}
\vspace{0.7cm}

\begin{singlespace}
\noindent \begin{center}
Volume 47B, Number 4 Physics Letters 6 August 1973
\par\end{center}

\noindent \begin{center}
\textbf{\Large Nuclear Compressibility}
\par\end{center}{\Large \par}

\noindent \begin{center}
L. Zamick 
\par\end{center}

\noindent \begin{center}
Rutgers University, New Brunswick, N.J., U.S.A
\par\end{center}

\noindent \begin{center}
Received 24 May 1973
\par\end{center}
\end{singlespace}
\begin{abstract}
A simple formula is obtained, which relates the nuclear compressibility
to the binding energy per particle, the mean kinetic energy per particle,
and the power of the density appearing in the two body interaction.
\end{abstract}
In this work we wish to calculate the compressibility of a nucleus,
assuming an interaction which compressibility of a nucleus, assuming
an interaction which correctly reproduces the total binding energy
and the mean square radius. Such that an interaction requires at least
two parameters in order to obtain these two properties, but we shall
use a three-parameter interaction \[
-\alpha\delta\left(r\right)+\gamma\rho^{\sigma}\left(R\right)\delta\left(r\right)\]
This a familiar density dependent interaction, not unsimilar to the
one chosen by Moszkowski {[}1{]}, except that he fixes the power of
the density, $\sigma$, to a value of 2/3 and he includes finite range
corrections. Vautherim and Brink {[}2{]} have $\sigma$=1. 

Here, on the contrary, the parameter $\sigma$ is kept as a variable,
for we wish to study what influence it has on the nuclear compressibility. 

We assume that we can restrict ourselves to harmonic-oscillator wave
functions, characterized by only one parameter, the oscillator length
b. In that case we can by dimensional analysis, write the expression
for the total binding energy as \[
E=-E_{B}=A/b^{2}+B/b^{3}+C/b^{\left(3+3\sigma\right)}\]
 where A, B, and C are independent of b. The first term is the kinetic
energy and is given by the oscilator formula \[
A/b^{2}=\left(\hbar/\left(2mb^{2}\right)\right)\sum2n+l+\nicefrac{3}{2}\]
Since we are at equilibrium \[
\frac{dE}{db}=0\]
\[
0=-2A/b^{3}-3B/b^{4}-\left(3+3\sigma\right)C/b^{\left(4+3\sigma\right)}\]
 We can regard the above as two equations in the two unknowns B and
C. We obtain \[
B=-b^{3}\left(1+\sigma\right)E_{B}/\sigma-\left\langle T\right\rangle b^{3}\left(1+3\sigma\right)/3\sigma\]
\[
C=b^{\left(3+3\sigma\right)}E_{B}/\sigma+\left\langle T\right\rangle b^{\left(3+3\sigma\right)}/3\sigma\]
 evaluated at the equilibrium value of b and $E{}_{B}$.

The nuclear compressibility K, is defined as\[
K=\left(1/\mathit{sA}\right)b^{2}\frac{d^{2}E}{db^{2}}\]
 We thus obtain an expression for K totally in terms of the binding
energy per particle and mean kinetic energy of a particle (both positive
quantities)\[
K=\left(1/\mathit{sA}\right)\left[\left\langle T\right\rangle +qE_{B}+\sigma\left(3\left\langle T\right\rangle +qE_{B}\right)\right]\]

This is our main result. Note that the compressibility is linear in
the power of the density $\sigma$ and increases with increasing $\sigma$. 

The above is an obvious explanation of Bethe's observation {[}3{]}
that the Vautherim Brink interaction {[}2{]} ($\sigma$=1), yields
a considerably larger compressibility then the interaction he used. 

\vspace{0.7cm}

\begin{center}
Volume 47B, Number 2 Physics Letters 29 October 1973
\par\end{center}

\begin{center}
\textbf{\Large Mass Parameter of the Breathing Mode State}
\par\end{center}{\Large \par}

\begin{center}
L. Zamick{*} 
\par\end{center}

\begin{center}
Institut fur Kernphysik, Kernforschunganlage Julich, D-517 Julich,
West Germany
\par\end{center}

\begin{center}
Received 22 August 1973
\par\end{center}
\begin{abstract}
A simple identity is used to show that when harmonic oscillator wave
functions are used, Inghs formula for the mass parameter of the breathing
mode state yields the classical result.
\end{abstract}
We note the following relationship obey by the operator $r{}^{2}$
(square radius) when Slater determinants with harmonic oscillator
radial wave functions are used\[
\sum_{n}\left|\left\langle 0\left|\sum r^{2}\right|n\right\rangle \right|^{2}=b^{2}\left\langle 0\left|\sum r^{2}\right|0\right\rangle \]
In the above b is the oscillator length parameter, |0) is the ground
state of a closed shell and |n) is an excited state. We need only
consider excited states of the type$\left|\alpha_{n+1}^{j+1}\alpha_{n}^{j}\right|^{J=0}$,
that is one-particle-one hole states which differ by a node and couple
to J=0. Working out both sides we get

\[
\sum\left(sl+1\right)\left(n+1\right)\left(n+l+3/2\right)=\sum\left(2l+1\right)\left(2n+l+3/2\right)\]

\[
n=0,1,2...\]

On the left we sum only over those occupied states nlj which are such
that the state n+1, lj is above the Fermi sea; on the right we sum
over all occupied states. 

The above can be derived by noting that the harmonic oscillator spectrum
is such that $(E_{n}-E_{o})$, the energy difference, is constant
$-2\hbar\omega$. Hence the energy independent sum rule for E(0) transitions
is proportional to the energy weighted sum rule. 

As noted by Werntz and Uberalll {[}2{]}, if one considers classically
a sphere of mass M and root mean square radius $\eta$ which undergoes
breathing mode oscillations then the kinetic energy is 1\textfractionsolidus{}2
$M\eta^{2}$. We wish to show that the same result is obtained in
the Inglis model {[}1{]} using harmonic oscillator wave functions. 

Let us first define an auxiliary mass parameter\[
M_{b}=2\hbar^{2}\sum_{n}\left|\left\langle n\left|\frac{\partial}{\partial b}\right|0\right\rangle \right|^{2}/\left(E_{f}-E_{i}\right)\]
 A harmonic oscillator wave function has the structure N(b)f(r/b),
where N is the normalization, and b is the oscillator length. We have\[
\frac{\partial}{\partial b}N\left(b\right)f\left(r/b\right)=\frac{dN}{db}f-r/bN\frac{\partial f}{\partial r}\]
 The first term does not contribute because of the orthogonality of
|n) and |0). Hence we get\[
M_{b}=\left(2\hbar^{2}\right)/b^{2}\sum_{n}\left|n\left\langle \left|\sum_{i}r_{i}\frac{\partial}{\partial r_{i}}\right|\right\rangle 0\right|^{2}\]
 The quantity $r\frac{\partial}{\partial r}$ can be written as a
commutator involving the shell model Hamiltonian\[
r\frac{\partial}{\partial r}=\left(-m/2\hbar^{2}\right)\left[H_{sm},r^{2}\right]-3/2\]
 he last term does not contribute. We also can make the simplification
$\left\langle n\left[H_{sm},r^{2}\right]0\right\rangle =\left(E_{f}-E_{i}\right)\left\langle n\left|r^{2}\right|0\right\rangle $.
Finally, but replacing $(E_{f}-E_{i})$ by $2\hbar\omega=\left(2\hbar^{2}\right)/\left(mb^{2}\right)$
we get\[
M_{b}=2\hbar^{2}\sum_{n}\left|\left\langle n\left|\frac{\partial}{\partial b}\right|0\right\rangle \right|^{2}/b^{4}\]

But we are interested in the mass parameter $M_{rms}$, in which the
root mean square radius $\eta$ is the dynamical variable. We get\[
M_{rms}=\left(\frac{\partial b}{\partial\eta}\right)^{2}M_{b}=\left(b^{2}M_{b}\right)/\left(\left\langle 0\left|\sum r^{2}\right|0\right\rangle /A\right)\]
 Hence, we get\[
M_{rmA}=mA\left\{ \frac{\sum_{b}\left|\left\langle n\left|\sum r^{2}\right|0\right\rangle \right|^{2}}{b^{2}\left(0\left|\sum r^{2}\right|0\right)}\right\} \]
 But from the relationship at the beginning of this paper the quantity
in brackets is one. Hence $M_{rms}=mA$. 

For completeness we note the well known equation of motion for the
breathing mode state\[
\frac{1}{2}M_{rms}\dot{\eta}^{2}+\frac{1}{2}AK\left(\eta-\eta_{o}\right)^{2}/\eta_{o}^{2}=E\]
 where K is the nuclear compressibility. The energy of the breathing
mode state is\[
\left(\hbar\omega\right)_{B}=\left(AK/\left(M_{rms}\eta_{o}^{2}\right)\right)^{1/2}=\sqrt{K/\left(m\eta_{o}^{2}\right)}\]

One often takes $\eta_{o}^{2}=\frac{3}{5}r_{o}^{2}A^{2/3}$ with $r_{o}$=1.2
fm. I would like to thank J. Speth and A. Fassler for their interest
and hospitality.

References

{[}1{]} D Inglis, Phys. Rev 97(1955) 701 

{[}2{]} C Werntz and H Überall, Phys. Rev 149(1966) 762 

\vspace{0.7cm}

\noindent \begin{center}
\textbf{\Large Isospin Mixing of the 1+ States in 12C{*}}
\par\end{center}{\Large \par}

\noindent \begin{center}
Hiroshi SATO and Larry Zamick 
\par\end{center}

\noindent \begin{center}
Department of Physics, Serin Physics Laboratory, Frelinghuysen Road,
Piscataway, New Jersey 08854, USA
\par\end{center}

\noindent \begin{center}
Received 26 July 1977
\par\end{center}
\begin{abstract}
The isospin mixing of the lowest two 1{*} states in 12C is here considered.
A nuclear charge symmetry breaking interaction (CSB) which was previously
fitted to help explain the Nolen-Schiffer anomaly, goes in the right
direction for this problem, but is not as important as in the N-S
problem. An important effect is mixing due to the one body Coulomb
field. This is shown to be approximately proportional to $\left(r^{2}\right)_{P_{1/2}}-\left(r^{2}\right)_{P_{3/2}}$,
a quantity which is very sensitive to the details of the one body
nuclear potential.
\end{abstract}
The total off-diagonal matrix elements are 170 keV with the DME and
92 keV with the SKII interaction. The DME can reproduce the experimental
value obtained by Lind et al., while the SKII interaction reproduces
the one obtained by Adelberger et al. 

Why are the DME and SKII interactions so different? Looking in table
2 we note that they differ by 62 keV in the off-diagonal matrix element
of one body coulomb field. Thus the quantity which is conceptually
the simplest thing entering into this calculation is giving the largest
ambiguity. 

By following an approximate procedure of Lane Martorelll and Zamick
{[}10{]}, we can relate the matrix element of the Coulomb potential
to the square radius of a shell model orbit. We approximate Uc by
its value in the interior of a uniform sphere\[
U_{c}\approx\frac{3Ze^{2}}{2R}\left(1-\frac{r^{2}}{3R^{2}}\right)\qquad\left(6\right)\]

Thus\[
\left\langle T=1\left|U_{c}\right|T=0\right\rangle =\frac{Ze^{2}}{4R^{3}}\left[\left\langle r^{2}\right\rangle _{\pi p_{1/2}}-\left\langle r^{2}\right\rangle _{\pi p_{3/2}}\right]\qquad\left(7\right)\]

We now see why we are getting large differences between DME and SKII
interaction. Whereas for the $p_{3/2}$ proton orbit the difference$\left\langle r^{2}\right\rangle _{DME}-\left\langle r^{2}\right\rangle _{SKII}-0.126\: fm^{2}$
or 1.7\%, the corresponding difference for the $p_{1/2}$ proton orbit
is much bigger, 0.962 fm$^{2}$ or 10\%. 

Why is there such a large difference between DME and SKII? A possible
explanation is that the $p_{1/2}$ orbit is bound by only 5.74 MeV
for DME, but for SKII interaction the value is 7.15 MeV. This certainly
goes in the right direction. 

We examine this possibility in more detail by making the following
list. Define$\Delta=\left[\left\langle r^{2}\right\rangle _{p_{1/2}}-\left\langle r^{2}\right\rangle _{p_{3/2}}\right]_{proton}$
\[
\Delta\left[fm^{2}\right]\left\langle T=1\left|U_{c}\right|T=0\right\rangle =\frac{Ze^{2}\Delta}{4R^{3}}\left[keV\right]\]

$\begin{array}{ccc}
DME & 2.172 & 178\\
WS[DME] & 1.844 & 151\\
SKII & 1.336 & 109\\
WS[SKII] & 1.347 & 109\end{array}$

In the above WS{[}DME{]} stands for the Woods-Saxon parameters which
give the same single particle energies as DME etc. Note that $\Delta$
is the same for the SKII interaction as for WS{[}SKII{]}. However,
there is a discrepancy between DME WS {[}DME{]} such that when all
is said and done the DME will give a larger off-diagonal matrix element.
Perhaps the most interesting comment from the above list is that we
get a change of 45 keV in going from WS {[}DME{]} to WS {[}SKII{]}.
This is, by simply demanding the $p{}_{1/2}$ orbit be bound by 5.74
MeV rather than 7.15 MeV, we are getting a change which is starting
to be comparable to the difference in the values which have been cited
by the two experimental groups. 

In conclusion we note that our results are not inconsistent with the
currently available experimental data; however, those data have large
error bars and there is some disagreement between two groups. The
phenomenological charge symmetry breaking interaction which helps
explain the Nolen-Schiffer anomaly also helps to enhance the isospin
mixing matrix element. The results depend sensitively upon the square
radii of the single particle orbits. Indeed two different Hartree-Fock
models give different results for these quantities, and it is hard
at present to favor one model over the other. Further work along these
lines i.e. learning what are the correct single particle energies
for the shell model orbits, is suggested.

References

{[}1{]} F.C. Barker, Nucl. Phys. 83(1966) 418. 

{[}2{]} J.M. Lind, G.T. Garvey and R.E. Tribble, Nucl. Phys. A276
(1977) 484. 

{[}3{]} E.G. Adelberger, R.E. Marrs, K.A. Snover and J.E. Bussoletti,
Phys, Lett. 62B (1976) 29, and Phys. Rev. C15 (1977) 484. 

{[}4{]} F.D. Reisman, P.I. Connors and J.B. Marion, Nucl. Phys. A153
(1970) 244. 

{[}5{]} H. Sato, Nucl. Phys. A269 (1976) 378. 

{[}6{]} D. Vautherin and D.M. Brink, Phys. Rev. C5 (1972) 626, and
J.W. Negele and D. Vautherin, Phys. Rev. C5 (1972) 1472, and also
unpublished. 

{[}7{]} G.F. Bertsch, and S. Shlomo, Phys. Rev. C10 (1974) 931. 

{[}8{]} E.H. Auerbach, S. Kahana, C.K. Scott and J. Weneser, Phys.
Rev. 188 (1969) 1747. 

{[}9{]} J.P. Schiffer and W.W. True, Rev. Mod. Phys. 48 (1976) 191. 

\vspace{0.7cm}

\noindent \begin{center}
PHYSICAL REVIEW C Volume 22, Number 4 October 1980
\par\end{center}

\noindent \begin{center}
\textbf{\Large Nuclear vibrations with a zero-range interaction and
the multipole condition}
\par\end{center}{\Large \par}

\noindent \begin{center}
Afsar Abbas And Larry Zamick 
\par\end{center}

\noindent \begin{center}
Serin Physics Laboratory, Rutgers University, Busch Campus, Piscataway,
New Jersey 08854 
\par\end{center}

\noindent \begin{center}
(received 15 January 1980)
\par\end{center}
\begin{abstract}
Isoscalar monopole, quadrupole, and octopole states are calculated
in closed shell nuclei. A delta interaction is used, its strength
determined by the multipole condition, namely that the mean single-particle-single-hole
potential energy difference is equal to the corresponding kinetic
energy difference. This interaction is used to obtain particle-hole
matrix elements appropriate for a random-phase approximation calculation.
The strengths given by different multipole conditions are different
but (except L=0) they appear to approach each other as the mass number
A becomes large. The monopole mode has already collapsed somewhat
before the strength implied by the quadrupole condition is reached.
Some of the calculations were repeated using a zero-range Skyrme interaction.
We observe a very high degeneracy in our calculation which we are
able to explain in terms of L-S coupling and the fact that the particle-hole
matrix element of a delta interaction is of the form$\int f_{P_{A}H_{A}}\left(r\right)f_{P_{B}H_{B}}\left(r\right)r^{2}dr$. 
\end{abstract}

\subsection*{IV. EXPLANATION OF HIGH DEGENERACY IN L-S }

Coupling Although a j-j coupling basis was used for these calculations,
it should be noted that the interaction we use is central, and we
are not introducing a one body spin orbit interaction. This means
that the quantum numbers L and S are good. This manifests itself in
the fact that most of the states that we obtain in an RPA diagonalization
have vanishing B(EL)'s to the ground state. These correspond to states
with S\ensuremath{\neq}0.

Furthermore, if we choose the single particle energies to be those
of a harmonic oscillator (this was described in more detail earlier
as SPII), we find that even fewer states than are permitted by L-S
coupling have finite B(EL) transition to ground states.

Furthermore, we observe that these states which have vanishing B(EL)'s
have eigenvalues which are some multiples of $\hbar\omega$. For the
quadrupole states there is a high degeneracy of $2\hbar\omega$; in
the octopole case at $1\hbar\omega$ and $3\hbar\omega$. Take for
example L=2$^{+}$ states in $^{16}O$. We find that there are 4 states
which are degenerate at exactly $2\hbar\omega$. These states have
zero B(E2) strength to the ground state. This degeneracy can be partly
explained by looking at the expression for the particle-hole interaction
in L-S coupling which is given in Appendix A. The particle-hole states
forming the 2$^{+}$ states in $^{16}O$ can be divided into 3 classes.

$\begin{array}{ccccc}
Class(1) & L\neq2 & S=1 & \left(0p^{-1}0f\right)L=3 & \left(0p^{-1}1p\right)L=1\\
Class(2) & L=2 & S=0 & \left(0s^{-1}0d\right)\left(0p^{-1}0f\right) & \left(0p^{-1}1p\right)\\
Class(3) & L=2 & S=1 & \left(0s^{-1}0d\right)\left(0p^{-1}0f\right) & \left(0p^{-1}1p\right)\end{array}$

The 3j symbol \[
\begin{array}{ccc}
l_{h} & l_{p} & \bar{L}\\
0 & 0 & 0\end{array}\]

vanishes unless $l{}_{h}+l_{p}+L$ is even. Hence all the matrix elements
in class 1 will vanish. This explains two of the four states at $2\hbar\omega$.

We next note that for every matrix element in class 2 that vanishes
there must be a corresponding one in class3 that vanishes. This is
because the entire spin dependence is contained in the factor$\left(1-4\delta_{S},_{o}\delta_{T},_{o}+2x\delta_{S},_{o}-2x\delta_{T},_{o}\right)$.
Since we are setting x=0, this factor will be 1 for S=1 and -3 for
S=0. The particle-hole matrix elements in class 2 are -3 times those
in class 3.

Thus two of the four degenerate states are from class 1, one from
class 2, and one from class 3. The ones from class 2 and class 3 are
expected to have the same radial structure, differing only in the
spins. Only class 2 states can have finite B(E2)'s. This is because
the E2 operator $\sum\gamma^{2}y_{2}$ has no spin dependence and
therefore cannot connect spin one to spin zero. Thus, in this example
we are left with one nontrivial degeneracy (class 2) to explain, as
well as the vanishing B(E2).

To explain this, we note that the L-S coupling expression for the
particle-hole matrix element (given in Appendix A) has the structure\[
\left\langle \left[p'h'^{-1}\right]^{LST}V\left[ph^{-1}\right]^{LST}\right\rangle =\int f_{k}'\left(r\right)f_{k}\left(r\right)r^{2}dr\]
 where from here on we use the symbol k to designate ph.

Let us denote the class 2 state at $2\hbar\omega$ by $\psi$ and
expand it in terms of particle-hole components (we limit ourselves
to TDA states in this discussion). \[
\psi=\sum z_{k}\left|k\right\rangle \]
 We Expect \[
\left(\psi V_{ph}\psi\right)=0\]

This can be achieved by demanding \[
\sum_{k}f_{k}\left(r\right)z_{k}=0\]

Let us first consider the schematic approximation in which the integral$\int R_{n{}_{1}l_{1}}\left(r\right)R_{n_{2}l_{2}}\left(r\right)R_{n_{3}l_{3}}\left(r\right)R_{n_{4}l_{4}}\left(r\right)r^{2}dr$
is replaced by a constant.$^{4}$ In this case the particle hole matrix
element has the structure$\left\langle k'V_{ph}k\right\rangle =g_{k}'g_{k}$,
where g goes not depend on r. We can regard $g_{k}$ as elements of
a vector of dimension D, where D is the number of particle-hole states
for a given LST configuration, e.g., in the above example, $^{16}$O
L=2, S=0, T=0 we have D=3. The quantities $z_{k}$ are also elements
of a vector of dimension D, and the condition$\sum_{k}z_{k}g_{k}=0$
simply means that the vector $\left\{ z\right\} $ is orthogonal to
the vector $\left\{ g\right\} $.

There are clearly (D-1) vectors $\left\{ z\right\} $ which are orthogonal
to $\left\{ g\right\} $. Each of these (D-1) vectors will have a
vanishing particle-hole matrix element. Thus the class 2 degeneracy
at $2\hbar\omega$ will be (D-1). This will also be for the class
3 degeneracy. Thus for L=2, S-0, and T=0 states in $^{16}O$ we expect
a twofold degeneracy from class 2. This was confirmed by a calculation.
We obtain the same degeneracy in class 3.

We now consider the case where $f{}_{k}\left(r\right)$ is not approximated
by a constant. We note that $f{}_{k}\left(r\right)$ is proportional
to the product of two harmonic oscillator radial wave functions $R{}_{n}\left(r\right)$,
$R{}_{h}\left(r\right)$. Using the Variable x=r/b we note that the
product can be written in the form of an exponential times a polynomial\[
f_{ph}\left(r\right)=e^{-x^{2}}X^{\sigma}\sum_{N_{min}\left(ph\right)}^{N_{max}\left(ph\right)}a_{n}\left(ph\right)X^{2n}\]
 where $\sigma$=0 or 1. Let $N_{min}$=minimum of all $N_{min}(ph)$
and let $N_{max}$= maximum of all $N_{max}(ph)$. To ensure that
$\sum_{k}f_{k}\left(r\right)z_{k}=0$ we demand that each coefficient
of$X{}^{2n+\sigma}$ vanishes. That is, $\sum a_{n}\left(k\right)z_{k}=0$.
This leads to $\Delta$ conditions where $\Delta=N_{max}-N_{min}+1$.
We now have $\Delta$ vectors, the nth one of which is$\left[a_{n}\left(1\right),a_{n}\left(2\right),...a_{n}\left(D\right)\right]$.
The vector $\left\{ z\left(k\right)\right\} $ has to be orthogonal
to all of these. There are clearly $\left(D-\Delta\right)$ such vectors
$\left\{ z\left(k\right)\right\} $. Hence class 2 degeneracy is $\left(D-\Delta\right)$.

In our example (L=2, S=0, and T=0 in $^{16}$O) the polynomial has
terms in x$^{2}$and x$^{4}$. Thus $N_{min}$=1, $N_{max}$=2, and
hence $\Delta$=2. Thus the degeneracy $\left(D-\Delta\right)$ is
one.

In a major shell n, l, N (N=2n+1l) the lowest power that appears in
a polynomial is $x^{1}$; the highest power is $x^{N}$. Thus if the
particle is in the shell $n_{p}$, $l_{p}$, $N_{p}$, and the hole
is the shell $n_{h}$, $l_{h}$, $N_{h}$, then we have \[
N_{max}(ph)=\frac{1}{2}(N_{p}+N_{h})\; if\; N_{p}+N_{h}\; is\; even\]
 \[
N_{max}(ph)=\frac{1}{2}(N_{p}+N_{h}-1)\; if\; N_{p}+N_{h}\; is\; odd\]
 \[
N_{min}(ph)=\frac{1}{2}(l_{p}+l_{h})\; if\; l_{p}+l_{h}\; is\; even\]
 \[
N_{min}(ph)=\frac{1}{2}(l_{p}+l_{h}-1)\; if\; l_{p}+l_{h}\; is\; odd\]

Just to give another example, consider the L=2 and T=0 states in $^{40}$Ca.
There are seventeen states in all, five in class 1, six in class 2,
and six in class 3.

The entire degeneracy at $2\hbar\omega$ for the schematic model 15.
Five of the states are from class 1. Since D is 6 for class 2 the
degeneracy here is D-1=5. It is also 5 for class 3.

For the delta interaction the entire degeneracy at $2\hbar\omega$
is eleven. Five still come from class 1, leaving 3 from class 2 and
3 from class 3. We find that the quantity $\Delta$ is equal to three.
The class 2 degeneracy is therefore $D-\Delta=3$, as expected.

It is very easy to show that the B(EL) to ground is zero for these
degenerate states.

The B(EL) is proportional to $\int r^{L}\sum_{k}f_{k}\left(r\right)z_{k}dr$.
Since the integrand is zero, the integral will also be zero. The above
is not true when we approximate the radial integral by a constant.

It should be emphasized that while part of the above argument involved
spin isospin symmetry (as discussed many years ago by de Shalit and
Walecka$^{5}$), the crucial part involving class 2 degeneracies did
not.

It is also worthwhile noting that for L-S coupling the expression
for the particle-hole matrix element the spin-isospin factor is very
simple$\left(1-4\delta_{S},_{o}+2x\delta_{S},_{o}-2x\delta_{T},_{o}\right)$.
The values of this factor for the four different modes are -3 for
S=0, T=0; (1-2x) for S=1, T=0; (1+2x) for S=0, T=1; and 1 for S=1,
T=1. 

\vspace{0.7cm}

\noindent \begin{center}
Physics review C Volume 31, Number 5 May 1985
\par\end{center}

\noindent \begin{center}
\textbf{\Large Comparison of magnetic dipole excitations in the f7/2
shell region with the new collective in 156Gd}
\par\end{center}{\Large \par}

\noindent \begin{center}
L. Zamick 
\par\end{center}

\noindent \begin{center}
Department of Physics and Astronomy, Rutgers University Piscataway,
New Jersey 
\par\end{center}

\noindent \begin{center}
08854 (Received 8 November 1984) 
\par\end{center}
\begin{abstract}
It is noted that in a single j shell $f{}_{7/2}$ calculations in
the titanium isotopes one obtains M1 rates of about one single particle
unit in strength to states at about 4 MeV excitation. This is systematically
consistent with the recently discovered collective excitations in
$^{156}$Gd, and neighboring nuclei. In the single j shell case, through,
the spin and orbital contributions to the M1 matrix element are nearly
the same.
\end{abstract}
Recently a low lying magnetic dipole excitation has been discovered
in $^{156}$Gd at an excitation energy of 3.075 MeV with a strength$B\left(M1\right)\uparrow-\left(1.3\pm0.2\right)\mu_{N}^{2}$.$^{1}$
The experiment involving high resolution inelastic electron scattering
with the Darmstadt linear accelerator. This was followed by the discovery,
by the Darmstadt group, of low lying 1$^{+}$ states in other deformed
nuclei such as $^{154}$Sm, $^{154}$Gd, $^{164}$Dy, $^{168}$Er,
and $^{174}$Yb. 

The purpose of this Brief Report is to point 1 out that the presence
of An M1 state with an excitation energy of 3-4.5 MeV and the order
of 1 single particle unit of strength may be a more widespread phenomenon.
We demonstrate this by calculating M1 rates to the lowest 1$^{+}$
states of the even-even titanium isotopes, using the wave function
of McCullen, Bayman, and Zamick (MBZ)$^{14,\,15}$ and Ginocchio and
French.$^{16}$ 

In a single j shell one can replace the magnetic dipole operator $\mu$
by$\left(g_{p}L_{p}+g_{n}L_{n}\right)\mu_{N}$. where $L_{p}$ is
the angular momentum operator for the protons and $L_{n}$ for the
neutrons. We use the same quenched $g_{p}$ and $g_{n}$ as were used
originally,$^{34}$ $g_{p}-1.50$ and $g_{n}-0.39$. The M1 transition
must be proportional to $(g_{p}-g_{n})^{2}$, for otherwise, the total
angular momentum operator $J-L_{\nu}-L_{n}$ would be able to induce
an M1 transition for J-0 to J=1. This is clearly impossible. 

The MBZ wave functions$^{14,\,15}$ for the titanium isotopes are
of the form\[
\Psi^{J}=\sum_{L_{p},L_{n},\nu}D^{J}\left(L_{p},L_{n},\nu\right)\left[\left(f_{7/2}^{2}\right)_{\pi}^{L_{p}}\left(f_{7/2}^{n}\right)_{\nu}^{L_{n}}\right]^{J}\]
where $D^{J}(L_{p},L_{n})$ is the probability amplitude that two
protons couple to $L_{p}$ and n neutrons couple to $L_{n}$. The
parameter V is the seniority quantum number. This is relevant only
to the nucleir $^{46}$Ti, where for $L_{n}-2$ and 4 we have states
of both seniority 2 and seniority 4. 

The expression for the M1 transition is\[
B\left(M1\right)\uparrow=\left(3/4\pi\right)\mu_{N}^{2}\left(g_{p}-g_{n}\right)^{2}\times\left|\sum_{LV}D^{0}\left(L,LV\right)D^{1}\left(L,LV\right)\sqrt{\left(L\left(L+1\right)\right)}\right|^{2}\]
The quantity $(g_{p}-g_{n})$ is set equal to 1.89. The coefficients
$D^{-1}(L_{p},L_{n})$ are contained in Ref. 17. The results for the
energies and M1 rates are

\[
\begin{array}{ccc}
 & E(MeV) & B(M1)\uparrow\left(\mu_{N}^{2}\right)\\
^{44}Ti & 5.81 & 2.70\\
^{46}Ti & 4.00 & 1.70\\
^{48}Ti & 3.83 & 0.69\end{array}\]

In $^{44}$Ti we have equal numbers of neutrons and protons. All 1$^{+}$
states in the $f_{7/2}$ model have isospin T=1. The J=0$^{+}$ ground
state, of course, has T=0. This nucleus is therefore a special case. 

However, for $^{46}$Ti and $^{48}$Ti, the $1_{1}^{+}$ states have
the same isospin as the ground state. It is therefore not unreasonable
to compare these nuclei with $^{156}$Gd and its neighbors. Considering
the wide spread in mass number between the nuclei, the behavior is
remarkably similar. Perhaps then the presence of $1^{+}$ state with
single particle strength in the 3-4.5 MeV range is a widespread phenomenon. 

Whether the physical interpretation of the $1^{+}$ states in the
two different regions is the same is another question. Independent
of what the answer is, it will clearly be of interest to try to verify
the presence over a wide range of the periodic table of $1^{+}$ states
whose energies and strengths vary systematically with mass number. 

The expression for the M1 rate by Dieperink$^{7}$ is\[
B\left(M1\right)\uparrow=\frac{3}{4\pi}\left(4N_{\pi}N_{\nu}\right)/\left(\left(N_{\pi}+N_{\nu}\right)\left(\bar{g}_{\pi}-\bar{g}_{\nu}\right)^{2}\mu_{N}^{2}\right)\]
 where $N{}_{\pi}$ and $N{}_{\nu}$ are the number of proton and
neutron boson, respectively, and $\bar{g}_{\pi}$ and $\bar{g}_{\nu}$
are the boson g factors. In the single j shell model, the g factors
are the same for the boson pairs$\left(f_{7/2}^{2}\right)^{J=2}$
as for the single particle states. 

If we naively apply this formula to the $f_{7/2}$ shell, with $\left(\bar{g}_{\pi}-\bar{g}_{\nu}\right)$
set equal to $(g_{p}-g_{n})=1.89$ then we obtain for $^{46}$Ti.
\[
B(M1)\uparrow=2.27\mu_{N}^{2}\]
and for $^{48}$Ti. \[
B(M1)\uparrow=1.71\mu_{N}^{2}\]
 These are larger than what we calculate in the $f_{7/2}$ model. 

In the single j shell model, one cannot ascribe the collective mode
as a pure orbital mode. This is clear from the fact that we can replace
$g_{l}l+g_{l}s$ by $g_{j}j$, where for j-l+1/2 \[
g_{j}=(1\lyxmathsym{\textfractionsolidus}j)g_{j}+g_{s}\lyxmathsym{\textfractionsolidus}(2j)\]
 while for j-l-1/2 \[
g_{j}=[(l+1)\lyxmathsym{\textfractionsolidus}(j+1)]g_{l}-g_{s}\lyxmathsym{\textfractionsolidus}[2(j+1)]\]
If we use the free values of $g_{l}$ and $g_{s}$, then the orbital
contribution to $(g_{p}-g_{n})$ is 0.86, and the spin contribution
is 1.34. 

The common choices for the renormalized values are $g_{l}=1.1$ for
a proton and -0.1 for a neutron, and $g_{s}=0.7g_{s}$ (free) for
the isovector term. With these values, the orbital contribution is
1.03 and the spin part is 0.94. We see that the orbit and spin contributions
are nearly equal. 

There is one common bond with the IBA formula,$^{6,7}$ of course.
In both expressions one has a factor$\left(\bar{g}_{\pi}-\bar{g}_{\nu}\right)^{2}$. 

\vspace{0.7cm}

\noindent \begin{center}
PHYSICAL REVIEW C Volume 51, Number 3 March 1995
\par\end{center}

\noindent \begin{center}
\textbf{\Large Limited symmetry found by comparing calculated magnetic
dipole spin and orbital strengths in 4He}
\par\end{center}{\Large \par}

\noindent \begin{center}
M.S. Fayache and L. Zamick 
\par\end{center}

\noindent \begin{center}
Department of Physics and Astronomy, Rutgers University, Piscataway,
New Jersey 08855 
\par\end{center}

\noindent \begin{center}
(received 17 June 1994)
\par\end{center}
\begin{abstract}
Allowing for $2\hbar\omega$ admixtures in $^{4}$He we find that
the summed magnetic dipole isovector orbital and spin strengths are
equal. This indicates a symmetry which is associated with interchanging
the labels of the spin with those of the orbit. Where Higher admixtures
are included, the orbital sum becomes larger than the spin sum, but
the sums over the low energy region are still nearly the same.
\end{abstract}
In Table I we give the total summed strength $B(M1)_{spin}$ and $B(M1)_{orbit}$
to all (nonspurious) $J=1^{+}$, T=1 states corresponding to the operators
$\overrightarrow{s}t_{z}$ and $lt_{z}$, respectively {[}as mentioned
before we drop the isovector factor 5.586-(-3.826)=9.412{]}. We do
this for progressively increasing model spaces: up to $2\hbar\omega$,
up to $4\hbar\omega$, and up to $6\hbar\omega$. 

We perform the calculations with the spin-orbit and tensor interactions
off and on.

Examining Table I we find one priori unexpected result. When we restrict
the ground state correlations to $2\hbar\omega$, we find that the
summed spin strengths are virtually equal to the summed orbital strengths.
This is true for all four cases of (x,y), i.e., whether or not there
is a tensor interaction present.

Table I. Summed spin and orbital magnetic dipole moment strengths
in$^{4}$He in units of $10^{-3}\mu_{N}^{2}$. 

$\begin{array}{|cc|cc|cc|cc|}
\hline Inter & action & up\; to\;6\hbar\omega &  & up\; to\;4\hbar\omega &  & up\; to\;2\hbar\omega & \\
\hline x & y & Spin & Orbit & Spin & Orbit & Spin & Orbit\\
\hline 0 & 0 & 0.8546 & 0.8546 & 1.3357 & 5.1635 & 1.5897 & 7.1474\\
1 & 0 & 0.8569 & 0.8571 & 1.3417 & 5.1851 & 1.6211 & 7.2296\\
0 & 1 & 3.8245 & 3.8239 & 5.2346 & 10.937 & 6.0653 & 14.607\\
1 & 1 & 3.3944 & 3.3955 & 4.8288 & 10.554 & 5.6052 & 14.272\\\hline \end{array}$

TABLE II. For the cases x=0, y=0 (central interaction, LS limit),
we give the energies and B(M1)'s of \textquotedbl{}spin excited\textquotedbl{}
and \textquotedbl{}orbit excited\textquotedbl{} states, with up to
$2\hbar\omega$ admixtures. 

$\begin{array}{|c|c|cc|}
\hline  & Energy\;(MeV) & B(M1)\;(in & units\; of\;10^{-3}\mu_{N}^{2})\\
\hline  &  & Spin & Orbit\\
\hline Nonspurious & 3.67 & 0 & 0\\
 & 44.0 & 0 & 0\\
 & 45.3 & 0.855 & 0\\
 & 48.8 & 0 & 0\\
 & 49.2 & 0 & 0\\
 & 53.9 & 0 & 0\\
 & 56.5 & 0 & 0.835\\
\hline Spurious & 436.7 & 0 & 0\\
 & 436.7 & 0 & 0\\
 & 436.7 & 0 & 0\\
 & 439.3 & 0 & 13.07\\\hline \end{array}$

We consider the case x=y=0. We are in the LS limit. Since the 0s$^{4}$
closed shell has L=0, S=0, only $2\hbar\omega$ excitations with the
same quantum numbers will admix into the ground state. Let us consider
two particles excited from the 0s shell to the 0p shell. We can label
the 2p-2h states by$\left[L_{\pi}L_{\nu}\right]^{L=0}\left[S_{\pi}S_{\nu}\right]^{S=0}$.
There are several cases to be considered: 

(1) Two protons are excited. The configurations are $\left(p_{\pi}^{2}\right)^{L_{\pi}S_{\pi}}\left(s_{\nu}^{2}\right)^{L_{\nu}S_{\nu}}$.
Since $L{}_{\nu}$=0 and $S{}_{\nu}$=0 and L and S are zero, we must
have $L{}_{\pi}$=0 and $S{}_{\pi}$=0. So all in all we get the state
$\left|a\right\rangle =\left(p^{2}\right)^{L_{\pi}=0,S_{\pi}=0}\left(s^{2}\right)^{L_{\nu}=0,S_{\nu}=0}$. 

(2) Two neutrons are excited. By analogy, the configuration is$\left|b\right\rangle =\left(s^{2}\right)^{L_{\pi}=0,S_{\pi}=0}\left(p^{2}\right)^{L_{\nu}=0,S_{\nu}=0}$. 

(3) A neutron and a proton are excited from the s shell to the p shell.
The configuration is $[(sp)^{L_{\pi}S_{\pi}}(sp)^{L_{\nu}S_{\nu}}]^{L=0,S=0}$.
There are two possibilities: $|c)=[L_{\pi}=1,L_{\nu}=1]^{L=0}[S_{\pi}=0,S_{\nu}=0]^{S=0}$
and $|d)=[L_{\pi}=1,L_{\nu}=1]^{L=0}[S_{\pi}=1,S_{\nu}=1]^{S=0}$. 

We can form an isovector orbital excitation by applying the operation
$\overrightarrow{L_{\pi}}-\overrightarrow{L_{\nu}}$ to the J=0$^{+}$
ground state; likewise we can form an isovector spin excitation by
applying the operator $\overrightarrow{S_{\pi}}-\overrightarrow{S_{\nu}}$
to the J=0$^{+}$ ground state. When acting on the configurations
|a) or |b), the orbital operator $\overrightarrow{L_{\pi}}-\overrightarrow{L_{\nu}}$
gives zero; likewise the spin operator $\overrightarrow{S_{\pi}}-\overrightarrow{S_{\nu}}$.
That is, $(\overrightarrow{L_{\pi}}-\overrightarrow{L_{\nu}})|L_{\pi}=0,L_{\nu}=0)=0$. 

Let us skip to the state |d). Note that the orbital and spin quantum
numbers are the same $L_{\pi}=S_{\pi}=1$ and $L_{\nu}=S_{\nu}+1$.
This is enough to prove that, if this were the only state present,
we would have the result $B(M1)_{spin}=B(M1)_{orbit}t$. 

In more detail, $(\overrightarrow{L_{\pi}}-\overrightarrow{L_{\nu}})|d)=N[L_{\pi}=1,L_{\nu}=1]^{L=1}[S_{\pi}=1,S_{\nu}=1]^{S=0}$
and $(\overrightarrow{S_{\pi}}-\overrightarrow{S_{\nu}})|d)=N[L_{\pi}=1,L_{\nu}=1]^{L=0}[S_{\pi}=1,S_{\nu}=1]^{S=1}$. 

There is no reason why these states should be at the same energy and
indeed they are not, but the equality of the spin and orbital strengths,
provided the state |c) were not present, is obvious. However, the
presence of the state |c) apparently presents a problem. The isovector
spin operator $\overrightarrow{S_{\pi}}-\overrightarrow{S_{\nu}}$
will annihilate this state, whereas the isovector orbital operator
$(\overrightarrow{L_{\pi}}-\overrightarrow{L_{\nu}})$ creates the
state $[L_{\pi}=1,L_{\nu}=1]^{L=1}[S_{\pi}=0,S_{\nu}=0]^{S=0}$. There
should therefore be more orbital strength than spin strength. What
saves the day is that this transition is spurious. In the OXBASH program
{[}10{]} the spurious states are put very high in energy by adding
a large constant to the single-particle energies for the center of
mass motion. We added 100 MeV for each nucleon, thus putting the spurious
states in the vincinity of 400 MeV excitation energy. In Table III
we show the $2\hbar\omega$ x=0, y=0 calculation in which all the
1+, T=1 states are shown, both nonspurious and spurious, with the
values of $B(M1)_{spin}$ and $B(M1)_{orbit}$. 

\vspace{0.7cm}

\noindent \begin{center}
PHYSICAL REVIEW C Volume 55, Number 3 March 1997
\par\end{center}

\noindent \begin{center}
\textbf{\Large Single-particle energies and Elliott's SU(3) model}
\par\end{center}{\Large \par}

\noindent \begin{center}
M. S. Fayache,$^{1}$ Y. Y. Sharon,$^{2}$ and L. Zamick,$^{3}$
\par\end{center}

\noindent \begin{center}
$^{1}$Départment de Physique, Faculté des Sciences de Tunis, Tunis
1060, Tunisia 
\par\end{center}

\noindent \begin{center}
$^{2}$Department of Physics and Astronomy, Rutgers University, Piscataway,
New Jersey 08855 
\par\end{center}

\noindent \begin{center}
$^{3}$TRIUMF, 4004 Wesbrook Mall, Vamcouver, British Columbia, Canada
V6T 2A3 
\par\end{center}

\noindent \begin{center}
(Received 6 August 1996)
\par\end{center}
\begin{abstract}
We address some properties of the quadrupole-quadrupole (Q\ensuremath{\centerdot}Q)
interaction in nuclear studies. We here consider how to restore SU(3)
symmetry even though we use only coordinate and not momentum terms.
Using the Hamilton $H=\sum_{i}\left(p^{2}/2m+m/2\omega_{i}^{2}r_{i}^{2}\right)\chi\sum_{i<j}Q\left(i\right)\cdot Q\left(j\right)-\left(\chi/2\right)\sum_{i}Q\left(i\right)\cdot Q\left(i\right)$
with $Q_{\mu}=r^{2}Y_{2,\mu}$, we find that only 2/3 of the single-particle
splitting $\epsilon_{0d}-\epsilon_{1s}$ comes from the diagonal term
of Q\ensuremath{\centerdot}Q; the remaining 1/3 comes from the interaction
of the valence nucleon with the core. The same is true in the 0f-1p
shell. {[}S0556-2813(97)01203-X{]}
\end{abstract}
Here we wish to obtain Elliott's SU(3) results {[}2{]} in a shell
model calculation in which only the coordinate Q\ensuremath{\centerdot}Q
interaction is used. We do not wish to used the momentum-dependent
terms. The latter were introduced by Elliott so that, in combination
with the coordinate terms, there would be no $\Delta$N=2 admixtures,
i.e., no admixture from configurations involving $2\hbar\omega$ excitations.
However, in many cases we want to see the effects of such admixtures
in our shell model studies. One classic problem in which $\Delta$N=2
admixtures are important is the isoscalar E2 effective charge which
gets enhanced by a factor of 2 when such admixtures are allowed. There
are many other problems of interest along these lines, some of which
we have considered {[}1{]}. 

The Hamiltonian we consider is therefore \[
H=\sum_{i}(\frac{p^{2}}{2m}+\frac{1}{2}m\omega^{2}r_{i}^{2})-\chi\sum_{i<j}Q(i)\centerdot Q(j)-\frac{\chi}{2}\sum_{i}Q(i)\centerdot Q(i)\]
where \[
Q(i){}^{k}\centerdot Q(j){}^{k}=(-1){}^{k}\lyxmathsym{\textsurd}(2k+1)r(i){}^{k}r(j){}^{k}[Y(i){}^{k}Y(j){}^{k}]^{0}\]
 with k=2. Like Elliott, we have not only the two-body Q\ensuremath{\centerdot}Q
term, but also the i=j single-particle term. 

It is convenient to introduce the quantity$\bar{\chi}=5b^{4}\chi/32\pi$
where b is the oscillator length parameter, such that $b{}^{2}=\hbar/m\omega=41.46/\hbar\omega$. 

To evaluate the single-particle term we use the addition theorem\[
p_{k}\left(\cos\theta_{12}\right)=\frac{4\pi}{2k+1}\sum_{\mu}Y_{k,\mu}\left(1\right)Y_{k,\mu}^{*}\left(2\right)\]

and thus\[
\sqrt{5}\left[Y^{2}\left(i\right)Y^{2}\left(i\right)\right]^{0}=\frac{5}{4\pi}P_{2}\left(1\right)=\frac{5}{4\pi}\]

The single-particle potential is then\[
U\left(r\right)=-\frac{\chi}{2}Q\left(i\right)\cdot Q\left(i\right)=-4\bar{\chi}\left(\frac{r}{b}\right)^{4}\]

The expectation values of U(r)/$\chi$ for the single-particle states,
0s, 0p, 0d, 1s, 0f, and 1p are, respectively, -15, -35, -63, -75,
-99, and -119. What single-particle splitting$\epsilon_{0d}-\epsilon_{1s}$
is needed to get Elliott's SU(3) results? The best way to answer this
is to give the formula for the SU(3) energy in the 1s-0d shell (in
which the momentum terms are included):\[
E\left(\lambda\mu\right)=\bar{\chi}\left[-4\left(\lambda^{2}+\mu^{2}+\lambda\mu+3\left(\lambda+\mu\right)\right)+3L\left(L+1\right)\right]\]
 For a rotational band, the L=2-L=0 splitting. 

The splitting due to the diagonal Q\ensuremath{\centerdot}Q interaction
is $\left[-63-\left(-75\right)\right]\chi=12\bar{\chi}$. This is
only 2/3 of the desired result. Where does the remaining $1/3\left(6\chi\right)$
come from? 

The answer is that the missing part comes from the interaction of
the valence particle with core. That is to say, in order to get Elliott's
SU(3) results we must not only include the diagonal term, but also
the particle-core interaction. 

The expression for the particle-core interaction is \[
\delta\epsilon_{j}=-\chi\sum_{c,m_{c}}\left\langle \Psi_{m}^{j}\left(1\right)\Psi_{m_{c}}^{c}\left(2\right)\left|Q\cdot Q\right|\Psi_{m}^{j}\left(1\right)\Psi_{m_{c}}^{c}\left(2\right)-\Psi_{m}^{j}\left(2\right)\Psi_{m_{c}}^{c}\left(1\right)\right\rangle \qquad\left(1\right)\]
where j and m represent all the quantum numbers of the valence nucleon
(including isospin labels which have been suppressed) and c and $m_{c}$
are the quantum numbers of a particle in the core. 

With the above Q\ensuremath{\centerdot}Q interaction only the exchange
term survives. The expression becomes\[
\delta\epsilon_{j}=\chi\sum_{c,m_{c}}\left\langle \Psi_{m}^{j}\Psi_{m_{c}}^{c}Q\Psi_{m_{c}}^{c}\Psi_{m}^{j}\right\rangle \qquad\left(2\right)\]

We obtain $\delta\epsilon_{0d}-\delta\epsilon_{1s}=6\bar{\chi}$.
This is missing 1/3 of the splitting required to get Elliott's SU(3)
results {[}1{]}. 

The above results for the 1s-0d shell are more general. In the 1p-0f
shell the single-particle splitting required to get the SU(3) result
is $\epsilon_{0f}-\epsilon_{1p}=3\left(3\times4=1\times2\right)\bar{\chi}=30\bar{\chi}$.
Once again we only get 2/3 of this $20\bar{\chi}$ from the diagonal
Q\ensuremath{\centerdot}Q term. The remaining $10\bar{\chi}$ comes
from the interaction of the valence nucleon with the core (actually
only the 0p shell in the core will contribute). 

\vspace{0.7cm}

\noindent \begin{center}
PHYSICAL REVIEW C Volume 56, Number 2 August 1997
\par\end{center}

\noindent \begin{center}
\textbf{\Large Need for an isovector quadrupole term in the sum rule
relating scissors mode excitations to B(E2) values}
\par\end{center}{\Large \par}

\noindent \begin{center}
Y.Y Sharon{*} and L. Zamick 
\par\end{center}

\noindent \begin{center}
Department of Physics, Rutgers University, Piscataway, New Jersey,
08855
\par\end{center}

\noindent \begin{center}
M.S. Fayache 
\par\end{center}

\noindent \begin{center}
Department de Physique, Faculté des Sciences de Tunis, Tunis 1060,
Tunisia
\par\end{center}

\noindent \begin{center}
G. Rosensteel 
\par\end{center}

\noindent \begin{center}
Department of Physics, Tulane University, New Orleans, Louisiana 70118 
\par\end{center}

\noindent \begin{center}
(Received 21 January 1997; revised manuscript received 14 March 1997)
\par\end{center}
\begin{abstract}
For a Q\ensuremath{\centerdot}Q interaction the energy-weighted sum
rule for isovector orbital magnetic dipole transitions is proportional
to the difference $\sum$B(E2, isoscalar)-$\sum$B(E2, isovector),
not just to $\sum$B(E2, physical). This fact is important in ensuring
that one gets the correct limit as one goes to nuclei, some of which
are far from stability, for which one shell (neuron or proton is closed.
{[}S0556-2813(97)03208-1{]}

PACS number(s): 21.60.Fw, 21.10.Re, 21.60.Ev, 21.60.Cs
\end{abstract}
Using the interaction $-\chi Q\cdot Q$, Zamick and Zheng {[}1,2{]}
obtained a sum rule which relates the scissors mode excitation rate
(i.e., the isovector orbital magnetic dipole excitation rate) to the
electric quadrupole excitation rate. The isovector orbital magnetic
dipole operator is $(\overrightarrow{L_{\pi}}-\overrightarrow{L_{\nu}})\lyxmathsym{\textfractionsolidus}2$
{[}the isoscalar one is half the total orbital angular momentum $\overrightarrow{L}\lyxmathsym{\textfractionsolidus}2=(\overrightarrow{L_{\pi}}-\overrightarrow{L_{\nu}})\lyxmathsym{\textfractionsolidus}2${]}.
In more detail, sum rule reads\[
\sum_{k}\left(E_{n}-E_{o}\right)B\left(M1\right)_{n,\uparrow}=\frac{9\chi}{16\pi}\sum_{i}\left\{ \left[B\left(E2,0_{1}\rightarrow2_{i}\right)_{IS}-B\left(E2,0_{1}\rightarrow2_{i}\right)_{IV}\right]\right\} \qquad\left(1\right)\]
where $B(M1)_{n,\uparrow}$ is the value for the isovector orbital
M1 operator $\left(g_{l\pi}=0.5,g_{l\nu}=-0.5,g_{s\pi}=0,g_{s\nu}=0\right)$
and the operator for the E2 transitions is$\sum_{protons}e_{p}r^{2}Y_{2}+\sum_{neutrons}e_{n}r^{2}Y_{2}$
with $e_{p}=1$, $e_{n}=1$ for the isoscalar transition(IV). The
above result also holds if we add a pairing interaction between like
particles, i.e., between two neutrons and two protons. Our main objectives
in this work are to clarify the role of the isovector B(E2) in the
above formula and to compare the fermion and boson model approaches
to scissors mode excitations. 

The above work was motivated by the realization from many sources
that there should be a relation between the scissors mode excitation
rate and nuclear collectivity. Indeed, the initial picture by LoIudice
and Palumbo {[}3{]} was of an excitation in a deformed nucleus in
which the symmetry axis of the neutrons vibrated against that of the
protons. In 1990-1991 contributions by the Darmstadt group {[}4,5{]},
it was noted that the Sm isotopes, which undergo large changes in
deformation as a function of mass number, the $B(M1)_{scissors}$,
was proportional to $B(Es,0_{1}\rightarrow2_{1})$. The B(E2) in turn
is proportional to the square of nuclear deformation , $\delta^{2}$. 

The above energy-weighted sum rule of Zamick and Zheng {[}2{]} was
attempt to obtain such a relationship microscopically using fermions
rather than interacting bosons. To a large extent they succeeded,
but there were some differences relative to {[}4,5{]}. Rather than
being proportional to $B(E2,0_{1}\rightarrow2_{1})$, the proportionality
factor was the difference in the summed isoscalar and summer isovector
B(E2)'s. Now one generally expects the isoscalar B(E2), especially
to the first $2^{+}$ state, to be the most collective and much larger
than the isovector B(E2). If the latter is negligible, then indeed
one basically has the same relation between scissors mode excitations
and nuclear collectivity, as empirically observed in the Sm isotopes. 

However, the derivation of the above energy-weighted sum rule is quite
general and should therefore hold (in the mathematical sense) in all
regions, not just where the deformation is strong. To best illustrate
the need for the isovector B(E2), consider a nucleus with a closed
shell of neutrons or protons. In such a nucleus, and neglecting ground
state correlations, the scissors mode excitation rate will vanish
as one needs both open shell neutrons and protons to get a finite
scissors mode excitation rate. However, if we have, say, an open shell
of protons to get a finite scissors mode excitation rate. However,
if we have, say an open shell of protons and a closed shell of neutrons,
the $B(E2,0_{1}\rightarrow2_{1})$ can be quite substantial. Many
vibrational nuclei are of such a type, and they have large B(E2)'s
from the ground state, e.g., 20 Weisskopf units (W.u.). 

However, in the above circumstances (closed neutron shell), the neutrons
will not contribute to the B(E2) even if we give them an effective
charge. But if only the protons contribute, it is clear that B(E2,
isovector)=B(E2, isoscalar). 

As an example, let us consider the even-even Be isotopes$^{6}$Be,$^{8}$Be,$^{10}$Be,
and$^{12}$Be. In doing so, we go far away from the valley of stability,
but this is in line with modern interests in radioactive beams.

BRIEF REPORTS 

TABLE I. The values of $B(M1)_{orbital},\; B(E2)_{isoscalar},\; B(E2)_{isovector}$
for Be isotopes. 

$\begin{array}{|c|c|c|c|c|c|}
\hline Nucleus &  & B(fm)^{a} & B(M1)_{orbital} & B(M1)_{isoscalar} & B(M1)_{isovector}\\
\hline ^{6}Be &  & 1.553 & 0 & 15.63^{b} & 15.63^{b}\\
^{8}Be &  & 1.597 & 0.637 & 73.54^{c} & 6.47\\
^{10}Be & T=1\rightarrow T=1 & 1.635 & 0.0895 & 69.7 & 31.27\\
 & T=1\rightarrow T=2 &  & 0.149 & 0 & 3.20\\
^{12}Be &  & 1.669 & 0 & 20.85 & 20.85\\\hline \end{array}$

$^{a}$b$^{2}$=41.46/($\hbar\omega$), $\hbar\omega=45/A^{2/3}-25/A^{1/3}$. 

$^{b}$The analytic expression in $^{6}$Be is $B(E2)=(50/4\pi)b^{4}e_{p}^{2}$. 

$^{c}$The analytic expression in $^{8}$Be is $B(E2)=(35/4\pi)b^{4}(ep+en)^{2}$.

Fayache, Sharma, and Zamick {[}6{]} have previously considered $^{8}$Be
and $^{10}$Be. The point was made that these two nuclei had about
the same calculated $B(E2,0_{1}\rightarrow2_{1})$, but the isovector
orbital B(M1)'s were significantly smaller in $^{10}$Be than in $^{8}$Be.
This went against the systematic that B(M1)orbital is proportional
merely to B(E2). In detail, the calculated $B(M1,0_{1}\rightarrow1)$
was $(2/\pi)\mu_{N}^{2}$ for $^{8}$Be and in $^{10}$Be was $(9/32\pi)\mu_{N}^{2}(T=1\rightarrow T=1)$
and $(15/32\pi)\mu_{N}^{2}(T=1\rightarrow T=2)$. Thus the ratio of
isovector orbital B(M1)'s is $^{10}Be/^{8}Be=3/8$. 

\vspace{0.7cm}

\noindent \begin{center}
Physical Review C 73, 044302 (2006)
\par\end{center}

\noindent \begin{center}
\textbf{\Large Seniority Conservation and Seniority Violation in the
g9/2 shell}
\par\end{center}{\Large \par}

\noindent \begin{center}
A. Escuderos and L. Zamick 
\par\end{center}

\noindent \begin{center}
Department of Physics and Astronomy, Rutgers University, Piscataway,
New Jersey 08854, USA 
\par\end{center}

\noindent \begin{center}
(Received 2 December 2005; published 6 April 2006)
\par\end{center}
\begin{abstract}
The g$_{9/2}$ shell of identical particle is the first one for which
one can have seniority-mixing effects. We consider three interactions:
a delta interaction that conserves seniority, a quadrupole-quadrupole
(Q\ensuremath{\centerdot}Q) interaction that does not, and a third
one consisting of two-body matrix elements taken from experiment ($^{98}$Cd)
that also leads to some seniority mixing. We deal with proton holes
relative to a Z=50, N=50 core. One surprising result is that, for
a four-particle system with total angular momentum I=4, there is one
state with seniority $\nu$=4 that is an eigenstate of any two-body
interaction-seniority conserving or not, The other two states are
mixtures of $\nu$=2 and $\nu$=4 for the seniority-mixing interactions.
The same thing holds true for I=6. Another point of interest is that,
in the single-j shell approximation, the splitting $\Delta E=E(I_{max})-E(I_{min})$
are the same for three and five particles with a seniority conserving
interaction (a well-known result), but are equal and opposite for
a Q\ensuremath{\centerdot}Q interactions. The Z=40, N=40 core plus
g$_{9/2}$ neutrons (Zr isotopes) is also considered, although it
is recognized that the core is deformed.
\end{abstract}

\subsubsection*{III. SPECIAL BEHAVIORS FOR I=4+ AND 6+ STATES OF THE g9/2 CONFIGURATION }

For a system of four identical nucleons in the g$_{9/2}$ shell, the
possible seniorities are $\nu$=0,2, and 4, with $\nu$=0 occurring
only for a state of total angular momentum I=0. There is also a $\nu$=4
state with I=0. 

For !=4 and 6, we can have three states, one with seniority $\nu$=2
and two with seniority $\nu$=4. For the two $\nu$=4 states we have
at hand, we can construct different sets of $\nu$=4 states by taking
linear combinations of the original ones. If the original ones are
(4)$_{1}$ and (4)$_{2}$, we can form \[
(4)_{A}=a(4)_{1}+b(4)_{2},\;\quad(2)\]
 \[
(4)_{B}=-b(4)_{1}+a(4)_{2}\]
with $a^{2}+b^{2}=1$. The set (4)A, (4)B is as valid as the original
set. 

However, we here note that if we perform a matrix diagonalization
with any two-body interaction-seniority conserving or not-one state
emerges which does not depend on what the interaction is. The other
two states are, in general, mixtures of $\nu$=2 and $\nu$=4 which
do depend on the interaction. The values of the coefficients of fractional
parentage (cfp's) of this unique state of seniority 4 as shown in
Table I. The states of $J_{o}\neq4.5$ all have seniority $\nu$=4
state there is no admixture of states with $J_{o}=j=9/2$, be they
$\nu$=1 or $\nu$=3. Again, no matter what two-body interaction is
used, this I=4 state remains a unique state.

TABLE I: A unique J=4, $\nu$=4 SFP for j=9/2 

$\begin{array}{|c|c|}
\hline J_{o} & \left(j^{3}J_{o}j|j^{4}\; I=4,\nu=4\right)\\
\hline 1.5 & 0.1222\\
2.5 & 0.0548\\
3.5 & 0.6170\\
4.5\;(\nu=1) & 0.0000\\
4.5\;(\nu=3) & 0.0000\\
5.5 & -0.4043\\
6.5 & -0.6148\\
7.5 & -0.1597\\
8.5 & 0.1853\\\hline \end{array}$

Amusingly, this state does not appear in the compilation of seniority-classified
cfp's of Bayman and Lande {[}20{]} or de Shalit and Talmi {[}5{]}.
We should emphasize that, although different, the Bayman-Lande cfp's
are perfectly correct (as are the ones of de Shalit and Talmi, whose
cfp's are also different from those of Bayman and Lande {[}20{]}).
But then, why do they not obtain the unique state that we have shown
above? Bayman and Lande use group theoretical techniques to obtain
the cfp's diagonalizing the following Casimir operator for Sp(2j+1):
\[
G(Sp_{2j+1})=\frac{1}{2j+1}\sum_{odd\; k=1}^{2j}(-1)^{k}(2k+1)^{3/2}[U^{k}U^{k}]_{0}^{0}\qquad(3)\]
 where $U_{q}^{k}\equiv\sum_{i=1}^{N}U_{q}^{k}(i)$ and U is the Racah
unit tensor operator\[
\left\langle \Psi_{m'}^{j'}\left|U_{q}^{k}\right|\Psi_{m}^{j}\right\rangle =\delta_{jj'}\left(kjqm|j'm'\right)\qquad\left(4\right)\]
  The two seniority $\nu$=4 states are degenerate with such an interaction
and, since there is no seniority mixing, we can have arbitrary linear
combinations of the $4^{+}$ states. Only by using an interaction
which removes the degeneracy and violates seniority, do we learn about
the special state in Table I.

\subsubsection*{IV. THE ENERGY SPLITTING E(I\_max )-E(I\_min ) WITH A Q\ensuremath{\centerdot}Q
INTERACTION }

A well-known result for identical particles in a single j shell is
that, if one uses a seniority-conserving interaction, then the relative
spectra of states of the same seniority are independent of the number
of particles {[}5-7{]}. Thus, for n=3 and n=5, the seniority $\nu$=3
states have the same relative spectrum; for n=2, 4, and 6, the seniority
$\nu$=2 states have the same spectrum. These results hold, in particular,
for the delta interaction used here. 

Now the Q\ensuremath{\centerdot}Q interaction does not conserve seniority
and the above results do not hold. However, we have noticed an interesting
result for n=3 and n=5. Consider the splitting $E(I_{max})-E(I_{min})$,
$\nu$=3, where for $g_{9/2}$, $I_{max}=21\lyxmathsym{\textfractionsolidus}2$
and $I_{min}=3\lyxmathsym{\textfractionsolidus}2$. For a seniority-conserving
interaction, $\Delta E(n=5)=\Delta E(n=3)$, where for a Q\ensuremath{\centerdot}Q
interaction, $\Delta E(n=5)=-\Delta E(n=3)$. This will be discussed
quantitatively later.

\subsubsection*{X. THE $E(I_{max})-E(I_{min})$ SPLITTING FOR n=3 AND n=5: $^{97}$AG
VERSUS $^{95}$RH AND $^{83}$ZR VERSUS $^{85}$ZR }

As mentioned in a previous section, the splitting $\Delta E=E(I_{max}=21\lyxmathsym{\textfractionsolidus}2^{+})-E(I_{min}=3\lyxmathsym{\textfractionsolidus}2^{+})$
is the same for three particles as it is for five particles (or three
holes and five holes) if one has a seniority- conversing interaction.
However, for a pure Q\ensuremath{\centerdot}Q interaction, we have
$\Delta E(n=5)=-\Delta E(n=3)$. 

Using the $V(^{98}Cd)$ interaction for $^{97}$Ag and $^{95}$Rh,
we find \[
\Delta E(n=3)=0.77058\; MeV\]
 \[
\Delta E(n=5)=0.87818\; MeV\]
 They are both positive, an indication that the seniority-conserving
delta interaction is much more important than the seniority-violating
Q\ensuremath{\centerdot}Q interaction. 

Talmi had previously concluded, from an analysis of h11/2 nuclei with
a closed shell of neutrons (N=82), that seniority conservation held
to a high degree {[}7,31{]}. 

Unfortunately, for the $g_{9/2}$ nuclei that we are here considering
($^{92}$Tc, $^{95}$Rh, $^{97}$Ag, as well as the zirconium isotopes
$^{83}$Zr, $^{85}$Zr, $^{87}$Zr), although the high spin states
including $I=21/2^{+}$ have been identified, the $I=3/2^{+}$ states
in $^{97}$Ag, $^{95}$Rh, and $^{93}$Tc, as well as for Zr isotopes. 

For $^{83}$Zr and $^{85}$Zr, with a fitted interaction (despite
misgivings of using a single j model space), we find for $\Delta E=E(I_{max})-E(I_{min})$.\[
\Delta E(^{93}Zr)=0.48742\; MeV\]
 \[
\Delta E(^{95}Zr)=-0.59355\; MeV\]
 They have opposite signs, which shows that for these fitted interactions
the Q\ensuremath{\centerdot}Q interaction is much more important for
this case-neutrons beyond a Z=40, N=40 core. 

But it should be emphasized that the $I=3/2^{+}$ state is not part
of the fit because it has not been identified experimentally. If more
levels were known in the Zr isotopes, and in particular the low spin
level $I=3/2^{+}$ (but also $5/2^{+}$ and $1/2^{+}$), then the
picture might change. We strongly urge that experimental work be done
on all the nuclei considered here in order to locate the missing states,
especially $I=3/2{}_{1}^{+}$ and also $5/2{}_{1}^{+}$.

{[}1{]} A. F. Lisetskiy, B. A. Brown, M. Horoi, and H. Grawe, Phys.
Rev. C70, 044314 (2004) 

{[}2{]} D. J. Rowe and G. Rosensteel, Phys. Rev. Lett. 87,172501 (2001) 

{[}3{]} G. Rosensteel and D. J. Rowe, Phys. Rev. C 67,014303 (2003) 

{[}4{]} L. Zamick and A. Escuderos, Ann. Phys. 321, 987 (2006) 

{[}5{]} A. deShalit and I. Talmi, Nuclear Shell Theory (Academic Press,
New York, 1963) 

{[}6{]} R. D. Lawson, Theory of the Nuclear Shell Model (Clarendon
Press, Oxford, 1980)

\vspace{0.7cm}

\noindent \begin{center}
Annals of Physics 321 (2006) 987-998 
\par\end{center}

\noindent \begin{center}
\textbf{\Large New relations for coefficients of fractional parentage-The
Redmond recursion formula with seniority}
\par\end{center}{\Large \par}

\noindent \begin{center}
L. Zamick, A. Escuderos 
\par\end{center}

\noindent \begin{center}
Department of Physics and Astronomy, Rutgers University, Piscataway,
NJ 08854, USA
\par\end{center}

\noindent \begin{center}
received 26 July 2005; accepted 19 August 2005 Available online 29
September 2005
\par\end{center}
\begin{abstract}
We find a relationship between coefficients of fractional parentage
(cfp) obtained on the one hand from the principal-parent method and
on the other hand from a seniority classification. We apply this to
the Redmond formula which relates n\textrightarrow{}n+1 cfp's to n-1\textrightarrow{}n
cfp's where the principal-parent classification is used. We transform
this to the seniority scheme. Our formula differs from the Redmond
formula in as much as we have a sum over the possible seniorities
for the n\textrightarrow{}n+1 cfp's, whereas Redmond has only one
term. We show that there are useful applications of both the principal-parent
and the seniority classification.
\end{abstract}
A recursion formula for cfp's due to Redmond {[}1{]} is presented
in the books of de Shalit and Talmi {[}2{]} on p. 528, and Talmi {[}3{]}
on p. 274. It can be written as follows 

\[
(n+1)\left[j^{n}\left(\alpha_{0}J_{0}\right)jJ|\}j^{n+1}\left[\alpha_{0}J_{0}\right]J\right]\left[j^{n}\left(\alpha_{1}J_{1}\right)jJ|\}j^{n+1}\left[\alpha_{0}J_{0}\right]J\right]=\]
$\delta_{\alpha_{1}\alpha_{0}}\delta_{J_{1}J_{0}}+n(-1)^{J_{0}+J_{1}}\lyxmathsym{\textsurd}((2J_{0}+1)(2J_{1}+1))\sum_{\alpha_{2}J_{2}}{\begin{array}{ccc}
J_{2} & j & J_{1}\\
J & j & J_{0}\end{array}}$ \[
\times\left[j^{n-1}\left(\alpha_{2}J_{2}\right)jJ_{0}|\}j^{n}\alpha_{0}J_{0}\right]\left[j^{n-1}\left(\alpha_{2}J_{2}\right)jJ_{1}|\}j^{n}\alpha_{1}J_{1}\right]\qquad(7)\]
In the above, square bracket designates the principal parent used
to calculate the cfp. Actually, the principal parent sometimes looses
its significance because in some cases more than one principal parent
can yield the same cfp. In tables of cfp's, the principal parent is
usually not listed. The quantities in parentheses $(\alpha_{o}J_{o})$
are listed. The cfp with $(\alpha_{o}J_{o})$ is the probability amplitude
that a system of (n+1) identical particles with quantum numbers $(\alpha_{o}J_{o})$
and a single nucleon.

\paragraph*{2. Relation between principal-parent cfp's and those in the seniority
scheme }

We here note a relationship between the overcomplete set of principal-parent
coefficients of fractional parentage and those with the seniority
classification 

\[
\left[j^{n}\left(\nu_{0}J_{0}\right)jJ|\}j^{n+1}\left[\nu_{0}J_{0}\right]J\right]\left[j^{n}\left(\nu_{1}J_{1}\right)jJ|\}j^{n+1}\left[\nu_{0}J_{0}\right]J\right]=\]

\[
\sum_{\nu}\left[j^{n}\left(\nu_{0}J_{0}\right)jJ|\}j^{n+1}J\nu\right]\left[j^{n}\left(\nu_{1}J_{1}\right)jJ|\}j^{n+1}J\nu\right]\qquad(8)\]
In the left-hand side above, the first principal parent is formed
by adding the (n+1)th nucleon to an n-nucleon to an n-nucleon antisymmetric
system with good seniority and angular momentum $(\nu_{o}J_{o})$,
then coupling the combined sytem to a total angular momentum J, and
then antisymmetrizing and normalizing the total wave function. On
the right-hand side, the sum over $\nu$ is a sum over all the possible
seniorities of the combined (n+1) system and, for a given seniority,
over all states with that seniority. 

A proof of the above result will be given in Appendix A. 
\end{document}